# SOS: Symmetry Operational Similarity

**Sang-Wook Cheong**

Rutgers Center for Emergent Materials and Department of Physics and Astronomy, Piscataway, NJ 08854, USA

Correspondence: S.-W. Cheong (email: sangc@physics.rutgers.edu)

**Abstract**

**Symmetry often governs condensed matter physics. The act of breaking symmetry spontaneously leads to phase transitions, and various observables or observable physical phenomena can be directly associated with broken symmetries. Examples include ferroelectric polarization, ferromagnetic magnetization, optical activities (including Faraday and magneto-optic Kerr rotations), second harmonic generation, photogalvanic effects, nonreciprocity, various Hall-effect-type transport properties, and multiferroicity. Herein, we propose that observable physical phenomena can occur when specimen constituents (i.e., lattice distortions or spin arrangements, in external fields or other environments, etc.) and measuring probes/quantities (i.e., propagating light, electrons or other particles in various polarization states, including vortex beams of light and electrons, bulk polarization or magnetization, etc.) share symmetry operational similarity (SOS) in relation to broken symmetries. In addition, quasi-equilibrium electronic transport processes such as diode-type transport effects, linear or circular photogalvanic effects, Hall-effect-type transport properties ((planar) Hall, Ettingshausen, Nernst, thermal Hall, spin Hall, and spin Nernst effects) can be understood in terms of symmetry operational systematics. The power of the SOS approach lies in providing simple and physically transparent views of otherwise unintuitive phenomena in complex materials. In turn, this approach can be leveraged to identify new materials that exhibit potentially desired properties as well as new phenomena in known materials.**

**Introduction**

Symmetry appears in all areas of physics. For example, the local SU(3)xSU(2)xU(1) gauge symmetry and its spontaneous symmetry breaking is the essence of the standard model encompassing quarks, the τ neutrino and the Higgs boson. The theoretical prediction and experimental confirmation of the existence of these elementary particles has been a scientific triumph in the last half century. Symmetry and its breaking are also quintessential in numerous physical phenomena in condensed matter such as the appearance of bulk polarization or magnetization through phase transitions, the generation of second harmonic light with strong light illumination, and various nonreciprocal diode effects, etc.. There are five important symmetries relevant to crystalline materials, namely translational, rotational, mirror reflection, space inversion and time reversal. These symmetries are not completely independent; for example, space inversion operation is equivalent to a 180° rotation about the vertical axis plus a mirror reflection about the horizontal mirror plane (e.g., **I**=**R**⊗**M** with the definition discussed below). Comprehensive theoretical descriptions and classifications of all crystalline materials and also the requirement for, for example, nonreciprocity in crystalline materials in terms of full symmetry analysis are well documented in literature [1-9], but they are often complex and material-specific. In this article, we will limit our discussion to the central, often one-dimensional (1D), cases that are pictorially succinct and intuitive, applicable to numerous different materials, and most relevant to observables or observable physical phenomena.

When the motion of an object in one direction is different from that in the opposite direction, it is called a nonreciprocal directional dichroism or simply a nonreciprocal effect [7-9]. The object can be an electron, a phonon, spin wave, or light in crystalline solids, and the best known example is that of nonreciprocal charge transport (i.e., diode) effects in $p$–$n$ junctions, where a built-in electric field ($E$) breaks the directional symmetry [10]. The polarization ($P$) of ferroelectrics such as $BiFeO_3$ or polar semiconductors such as BiTeBr can also act like the built-in electric field, so bulk diode (and photovoltaic) effects can be realized in these materials [11,12]. Certainly, both $E$ and $P$ are polar vectors, and behave identical under various symmetry operations – similar with the identical behavior of magnetic field ($H$) and magnetization ($M$). In addition to $p$–$n$ junctions, numerous technological devices such as optical isolators, spin current diodes or metamaterials do utilize nonreciprocal effects.

Multiferroics, where ferroelectric and magnetic orders coexist, has attracted an enormous attention in recent years because of the cross-coupling effects between magnetism and ferroelectricity, and the related possibility of controlling magnetism with an electric field (and vice versa) [13-15]. Magnetic order naturally breaks time reversal symmetry, and a magnetic lattice, combined with a crystallographic lattice, can have broken space inversion symmetry, leading to multiferroicity, called magnetism-driven ferroelectricity. Since space inversion and time reversal symmetries are simultaneously broken in multiferroics with magnetic order and electric polarization, multiferroics are often good candidates for nonreciprocal effects.

**Results and Discussion**
**Nonreciprocal Directional Dichroism**

Nonreciprocity directional dichroism can be often understood from the consideration of SOS with velocity vector. Velocity vector ($k$, linear momentum or wave vector), which is d$x$/d$t$ where $x$=displacement and $t$=time, changes its direction under space inversion as well as time reversal, so is associated with broken space inversion and time reversal symmetries. $k$ can be associated with the motion of any quasi-particles such as electrons, spin waves, phonons, and photons in a specimen or the motion of the specimen itself. Since here, we are dealing with 1D measuring probes/quantities, we will use the following notations for various symmetry operations with respect to 1D measuring probes/quantities. For nonreciprocity (and other discussions except multiferroicity), the measuring probes/quantities are related with velocity vectors. Thus, we define **R**=π (i.e. 2-fold) rotation operation with the rotation axis perpendicular to the $k$ direction, <u>**R**</u>=π (i.e. 2-fold) rotation operation with the rotation axis along the $k$ direction, **I**=space inversion, **M**=mirror with the mirror plan perpendicular to the $k$ direction, <u>**M**</u>= mirror operation with the mirror plane containing the $k$ direction, **T**=time reversal operation. (In the standard crystallographic notations, "**R**, <u>**R**</u>, **I**, **M**, <u>**M**</u>, and **T**" are "$C_{2\perp}$, $C_{2\parallel}$, $\bar{1}$, $m_\perp$, $m_\parallel$, $1'$, respectively.) Evidently, +$k$ becomes -$k$ by **R** symmetry operation, i.e. $k$ has broken **R** symmetry. Similarly, +$k$ becomes -$k$ by any of **I**, **M**, and **T** symmetry operations. In fact, {**R**,**I**,**M**,**T**} is the set of all broken symmetries of $k$. Interestingly, all specimen constituents in the left-hand side of Fig. 1 have also broken {**R**,**I**,**M**,**T**}, which can be proven straightforwardly, but is a highly non-trivial statement. In these circumstances, we say that these specimen constituents do have SOS (symmetry operational similarity, shown with "≈" symbol in Fig. 1) with $k$, lacking a better terminology. Note that since we consider the 1D nature of $k$, translational symmetry along the 1D direction is ignored, and additional broken symmetries in the left-hand-side specimen constituents, such as **R** (different from π or 2π rotation) in Fig. 1e, is also ignored. When $+k$ becomes $-k$ under a symmetry operation while a specimen constituent where quasi-particles are moving with $\pm k$ is invariant under the symmetry operation, then the experimental situation becomes reciprocal. On the other hand, when a specimen constituent has SOS with $k$, then there is no symmetry operation that can connect these two experimental situations: one specimen constituent with $+k$ and the identical specimen constituent with $-k$. Thus, the experimental situation can become nonreciprocal, even though the magnitude of nonreciprocal effects cannot be predicted. Thus, all of the left-hand-side specimen constituents of Fig. 1, which have SOS with $k$, can exhibit nonreciprocal effects. For example, spin waves or light propagation in structurally-chiral (screw-type) magnet with magnetic fields along the chiral axis should exhibit nonreciprocal effects (the so-called magneto-chiral effects), which corresponds to Fig. 1(b). Nonreciprocity still works when the chiral axis is rotated by 90 degree, which can be coined as a transverse magneto-chiral effect. A nonreciprocal spin wave effect was observed in cubic chiral $Cu_2OSeO_3$ for spin wave propagating along the magnetization direction, in agreement with the magneto-chiral scenario shown in Fig. 1(b) [16]. A similar magneto-chiral spin wave effect was theoretically calculated, and also experimentally observed in monoaxial chiral $Ba_3NbFe_3Si_2O_{14}$ (Fe langasite) when the chiral axis, magnetic field, and spin wave propagation direction are all aligned [17-18]. Another example is nonreciprocal spin wave propagation in helical (screw-type) spins in magnetic fields along the helical spin axis or in conical spins, which corresponds to Fig. 1(d), and was observed in erbium metal [7]. Note that spin helicity does have SOS with structural chirality: both have broken {**I**,**M**,**M**}. Fig. 1(h) describes a toroidal moment situation and corresponds to the nonreciprocal THz optical effect observed in, for example, polar-ferromagnetic $FeZnMo_3O_8$ with pyroelectric **P** along the $c$ axis, **H** in the ab plane, and light propagation along the 3$^{rd}$ direction (perpendicular to both **P** and **H**) [19,20]. Note that Fig. 1(a) depicts structural ferro-rotation with **P** and **M** (or in **E** and **H**), and structural chirality with broken space inversion and ferro-rotation without broken space inversion are discussed in detail in ref. [15].

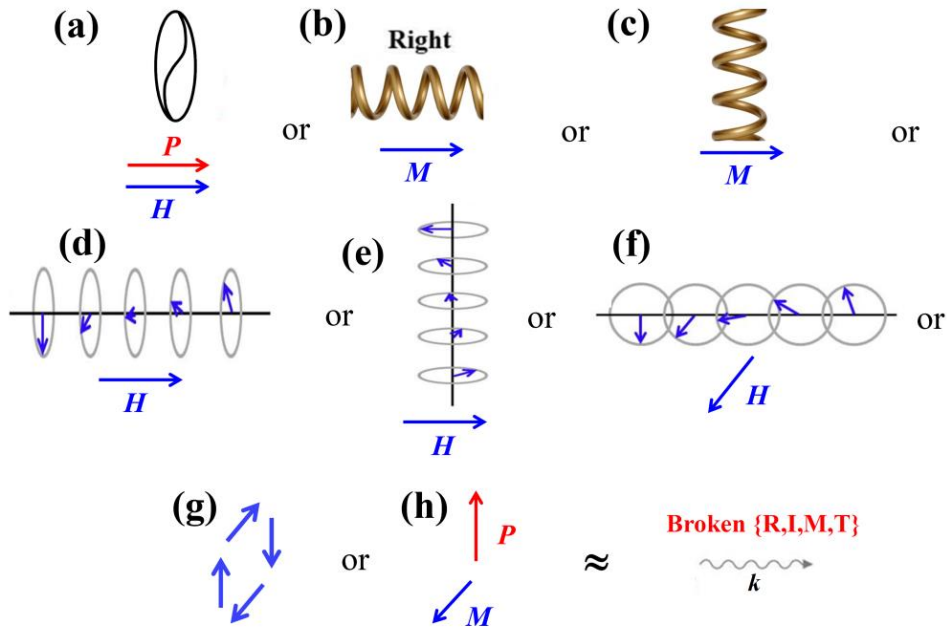

**Figure 1, Various specimen constituents having SOS with velocity vector.** The corresponding experiments should exhibit nonreciprocal directional dichroism. Red arrows: polarization (***P***) or external electric field (***E***), and blue arrows: magnetization (***M***) or external magnetic field (***H***). (a) ferro-rotation, (b) and (c) structural chirality (screw-type), (d) and (e) helical spin order (screw-type), (f) cycloidal spin order, (g) toroidal moment of rotating spins, and (h) toroidal moment with ***P*** and ***M*** (or ***E*** and ***H***).

The symmetry consideration that we discussed above also works more-or-less for quasi-equilibrium processes such as electronic transport experiments with ***k*** being the drift velocity of electron cloud, directly related with electric current vector ***J***. However, we need to add an external electric field to create ***k*** or ***J*** vector. Since {**R**,**I**,**M**} operations do link the left and right situations in Fig. 2, specimen constituents with electric polarization (or field), having broken {**R**,**I**,**M**}, as well as all nonreciprocal cases with broken {**R**,**I**,**M**}+{**T**} in Fig. 1 can exhibit nonreciprocal electronic transport effects. In fact, the diode effects in general *p-n* junctions and ferroelectric $BiFeO_3$ correspond to the nonreciprocal effect with ***P*** [10,11]. Magnetoresistance in chiral carbon nanotubes can be nonreciprocal, and corresponds to Fig. 1(b) – it is an electronic magneto-chiral effect [21]. When magnetic field is applied to BiTeBr in the *ab* plane with pyroelectric ***P*** along the *c* axis, electronic conductance along the 3$^{rd}$ direction (perpendicular to both ***P*** and ***H***) becomes nonreciprocal – i.e. exhibit an electronic nonreciprocal toroidal moment effect, corresponding to Fig. 1(h) [12].

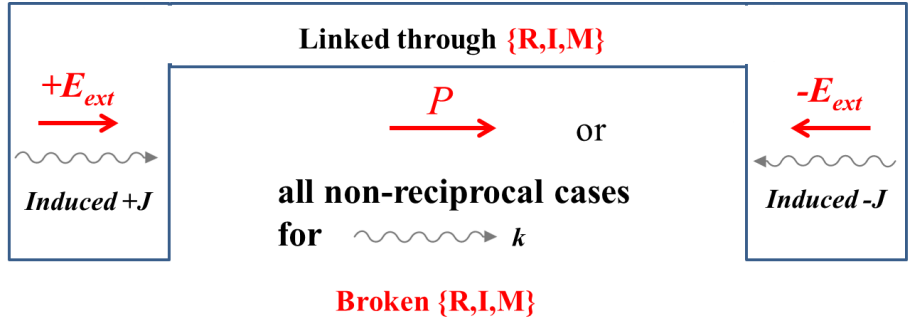

**Figure 2**, **Nonreciprocal electronic transport effects in quasi-equilibrium processes.** Two reciprocal situations can be linked through {**R**,**I**,**M**}, so materials with electric polarization (or field) with broken {**R**,**I**,**M**} as well as all nonreciprocal cases with broken {**R**,**I**,**M**}+{**T**} in Fig. 1 can exhibit nonreciprocal electronic transport effects.

**Multiferroicity and Linear Magnetoelectricity**

Multiferroicity, especially magnetism-driven ferroelectricity, can be also understood in terms of SOS with polarization (***P***) [22]. Since for multiferroicity, the measuring probes/quantities are ***P*** or magnetization (***M***), the symmetry operations of **R**, **R̲**, **M**, **M̲**, **I** and **T** for multiferroicity are defined with respect to the ***P***/***M*** direction, rather than the ***k*** direction. {**R**,**I**,**M**} is the set of all broken symmetries of ***P***, and all (quasi-)1D spin configurations in the left-hand side of Fig 3(a)-(d) have also broken {**R**,**I**,**M**}. Thus, they do have SOS with polarization. Note that since we consider the 1D nature of ***P***, translational symmetry along the 1D direction is irrelevant, and additional broken symmetries in the left-hand-side spin configurations, such as **R̲** (different from π or 2π rotation) in Fig. 3a, is also not relevant. Emphasize that ***P*** is not broken under the **T** operation, and the **T** symmetry is also not broken in all spin configurations in the left-hand side of Fig. 3a-d if proper translational operations are included. The cycloidal-spin-order-driven multiferroicity in, for example, TbMnO$_3$ and LiCu$_2$O$_2$ corresponds to Fig. 3(a) [23,24], and the multiferroicity driven by ferro-rotational lattice with helical spin order in, for example, RbFe(MoO$_4$)$_2$ and CaMn$_7$O$_{12}$ can be explained with Fig. 3(b) [25,26]. Fig. 3(c) corresponds to the multiferroicity with two (or more than two) different magnetic sites in Ca$_3$CoMnO$_6$, TbMn$_2$O$_5$ and orthorhombic HoMnO$_3$ [27-29]. The so-called *p-d* hybridization multiferroic Ba$_2$CoGe$_2$O$_7$ can be described by Fig. 3(d) [30]. The left-hand-side specimen constituents in Fig. 3(e) and (f) with zero *H* have only broken {**I**,**M**} and {**R**,**I**}, respectively, but they with non-zero *H* have now broken all of {**R**,**I**,**M**}, so becomes SOS with ***P***, which is consistent with, for example, the linear magnetoelectric effects in Cr$_2$O$_3$; diagonal linear magnetoelectric effect in low ***H***, corresponding to Fig. 3(e), and off-diagonal linear magnetoelectric effect in large ***H*** beyond the spin flop transition ***H***, corresponding to Fig. 3(f) [31]. Note that reversing the ***H*** direction in Fig. 3(e) and (f) (for example, by **R** and **M** operations, respectively) should accompanies the reversal of ***P***, which is consistent with the nature of linear magnetoelectricity. {**R**,**M**,**T**} is the set of all broken symmetries of ***M***. When ***H*** is replaced by electric field (***E***), the left-hand-side specimen constituents in Fig. 3(e) and (f) have broken {**R**,**M**,**T**}, so do have SOS with ***M***, which reflects the reciprocal relationship between electric field-induced magnetization and magnetic-field-induced polarization in linear magnetoelectrics such as Cr$_2$O$_3$.

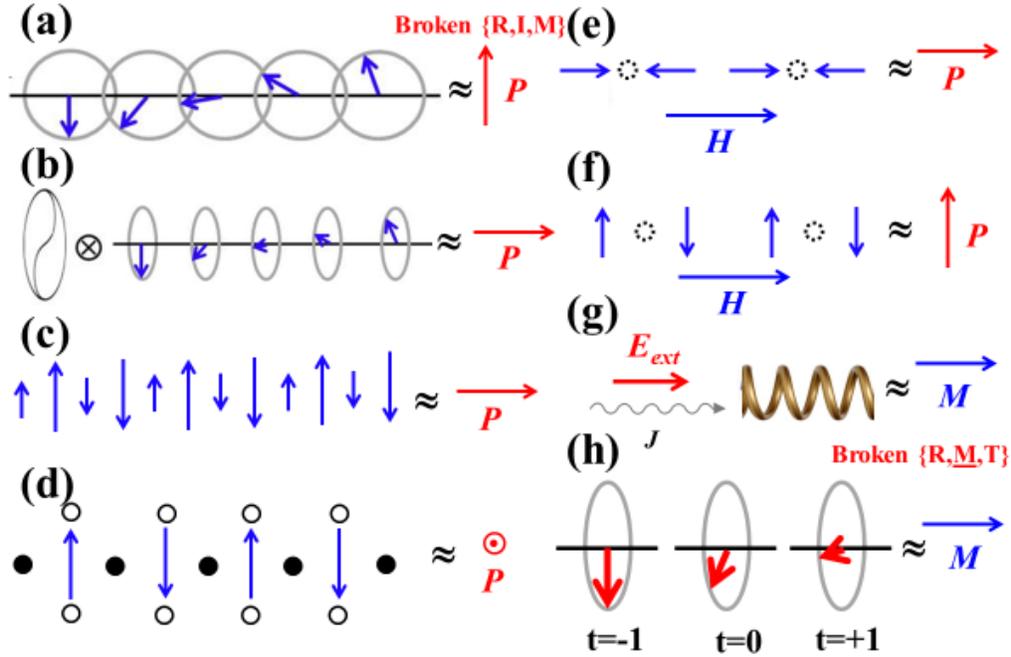

**Figure 3**, **Various (quasi-)1D specimen constituents having a SOS relationship with *P* or *M*.** Blue arrows are spins or *M*, and red arrows are *P*. (a) Cycloidal spin order. (b) Ferro-rotational lattice with helical spin order. (c) two-different alternating spins with up-up-down-down spin order. (d) Simple antiferromagnetic order with oxygens below (solid circles) and above (open circles) the page plane. (e) and (f) Simple antiferromagnetic order with alternating in-chain oxygens (dashed circles) in ***H***. (g) Current flow with external electric field in a screw-like chiral lattice. (h) Polarization rotating with time.

It is also interesting to consider how to induce magnetization in non-magnetic materials in non-trivial manners. Note that since any static configuration of ***P*** cannot break **T**, a time component has to be incorporated into the ***P*** configuration. Two structural examples having SOS with ***M*** are shown in Fig. 3(g) and (h). When electric current is applied to a tellurium crystal with a screw-type chiral lattice, corresponding to Fig 3(g), ***M*** can be induced in a linear fashion [32]. Consistently, Faraday rotation of linearly-polarized THz light propagating along the chiral axis of a tellurium crystal is also observed in the presence of electric current along the chiral axis, and this induced Faraday rotation effect is linearly proportional to the electric current [33]. Both of these highly-non-trivial effects correspond to the SOS relationship in Fig. 3(g). Fig. 3(h) can be realized when a Neel-type ferroelectric wall moves in the direction perpendicular to the wall [34]. This situation has not been experimentally realized yet, but the dynamic multiferroicity discussed in Ref. 35 is relevant to this case.

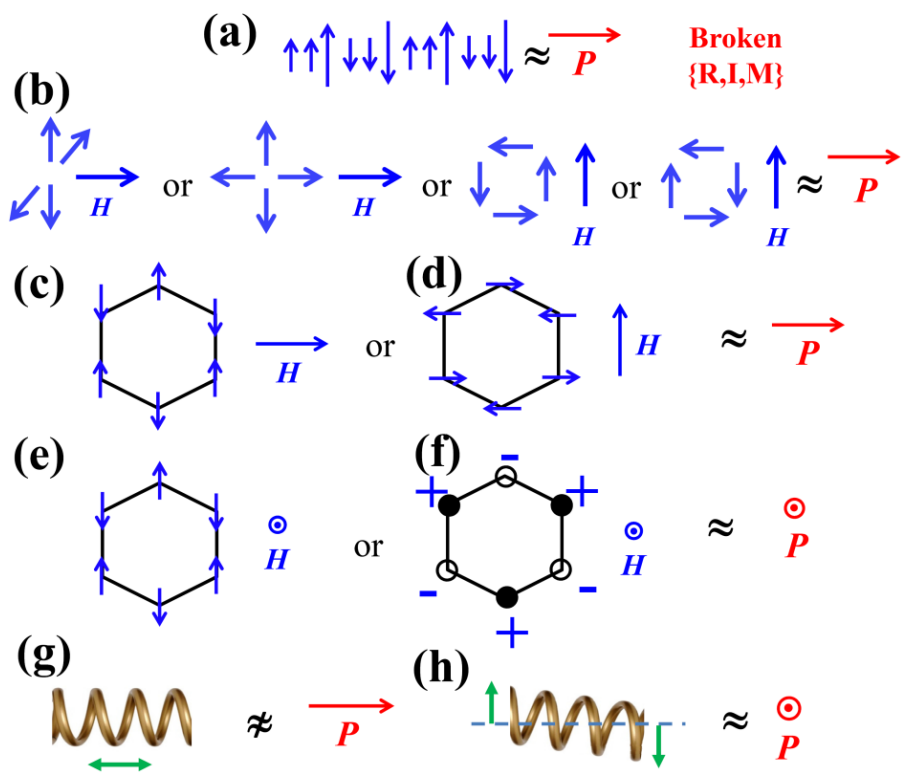

Figure 4, Various specimen constituents having a SOS relationship with *P*. Blue arrows are spins, and red arrows are *P*. (a) AAB-type ionic order combined with up-up-up-down-down-down spin order. (b) Magnetic monopole in perpendicular *H*, magnetic monopole in *H*, toroidal moment in *H*, and magnetic quadrupole in *H*. (c) – (e) Various in-plane antiferromagnetic order configurations on honeycomb lattice in *H* having SOS with *P*. (f) Buckled honeycomb lattice (spins below (solid circles) and above (open circles) the page plane) with Ising antiferromagnetic order in *H*, having SOS with *P*. Note that when the *H* direction in (f) is switched to, for example, the right direction in the page plane, then the specimen constituent has SOS with *P* to the right direction in the page plane. (g) Screw-type chirality under uniform strain (green double arrow; strain field). (h) Screw-type chirality under shear strain (two green arrows indicate shearing).

Our SOS approach for magnetism-driven ferroelectricity and linear magnetoelectricity, discussed above, is very powerful to predict new materials with magnetism-driven ferroelectricity and linear magnetoelectricity. For example, the combined configuration of ionic order with two types of magnetic ions (AAB type) and up-up-up-down-down-down-type magnetic order in Fig. 4(a) does have SOS with $P$, so is supposed to exhibit magnetism-driven ferroelectricity. The first two specimen constituents in Fig. 4(b) are for magnetic monopoles and the 3$^{rd}$ case in Fig. 4(b) is for magnetic toroidal moment, and the last specimen constituent in Fig. 4(b) is for magnetic quadrupole. No cases in Fig. 4(b) in zero $H$ has SOS with $P$, but all of them do have SOS with $P$ for non-zero $H$; thus, all can exhibit linear magnetoelectricity. We can also consider various spin configurations on (buckled) honeycomb lattice as shown in Fig. 4(c)-(f). None of the four cases in Fig. 4(c)-(f) in zero $H$ has SOS with $P$, but all of them in non-zero $H$ do have SOS with $P$, which indicates that all can be linear magnetoelectrics. Most of these cases have not been experimentally observed in real compounds, and it will be highly demanding to verify the symmetry-driven predictions in real materials [36]. One relevant example is hexagonal $R(Mn,Fe)O_3$ (R=rare earths), which is an improper ferroelectric with the simultaneous presence of Mn/Fe trimerization in the *ab* plane and ferroelectric polarization along the *c* axis. A number of different types of magnetic order with in-plane Mn spins have been identified in *h*-$R(Mn,Fe)O_3$. The so-called A1-type magnetic order in h-$R(Mn,Fe)O_3$ combined with Mn trimerization can induce a net toroidal moment, and the so-called A2-type magnetic order in h-$R(Mn,Fe)O_3$ combined with Mn trimerization can accompany a net magnetic monopole [37,38]. The linear magnetoelectric effect as well as nonreciprocity associated with these magnetic monopole and toroidal moment can be a topic for the future investigation.

**Piezoelectricity**

In general, piezoelectricity occurs in noncentrosymmetric crystallographic structures. However, the relevant 3x6 piezoelectric tensor is complex, and some of its components can be zero even in noncentrosymmetric systems. In fact, out of 32 point symmetry groups, 21 are noncentrosymmetric, but all of 3x6 piezoelectric tensor components in one (point group 432) of these 21 noncentrosymmetric groups are uniformly zero [1,2]. Here, we discuss how the presence of piezoelectricity can be understood in terms of our SOS concept. For piezoelectricity, specimen constituents consist of specimens under uniform strain or shear strain field, and we need to consider if a specimen constituent has a SOS relationship with an induced $P$. For example, Fig. 4(g) represents a screw-type chiral structure under uniform strain along any principle directions, which does not have SOS with $P$ in any direction, so piezoelectric $d_{11}=d_{22}=d_{33}$ coefficients should vanish (in fact, all of $d_{ij}$ (i,j=1,2,3) are zero). However, when shear strain field exists in a chiral structure in the page plane (see Fig. 4(h)), the specimen constituent does have SOS with $P$ perpendicular to the page plane, so $d_{14}$, $d_{25}$ and $d_{36}$ can be non-zero (other $d_{ij}$ (i=1,2,3 and j=4,5,6) are zero). This case corresponds to the piezoelectricity in, for example, point symmetry group $2_1 22$ (as well as 222).

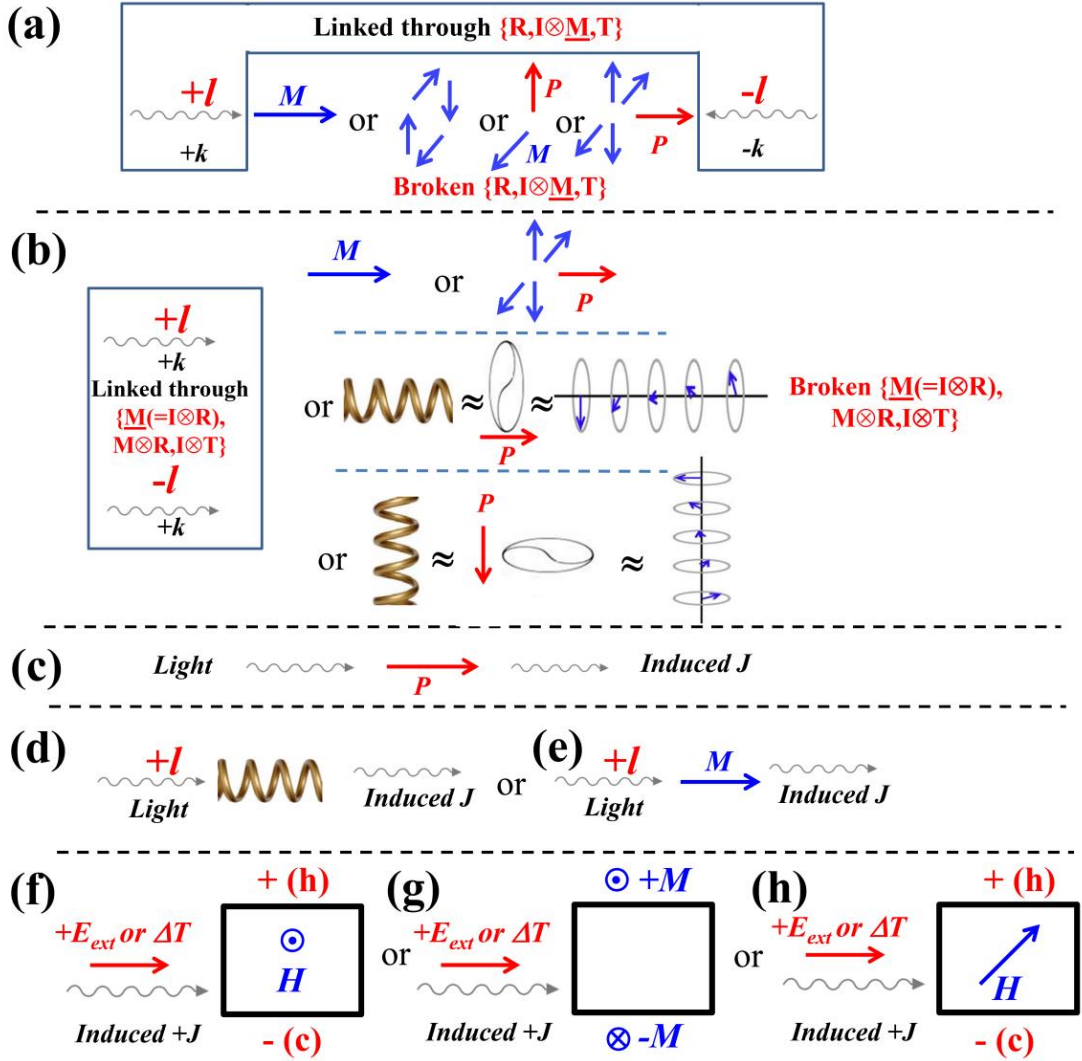

**Figure 5**, Various physical phenomena with the motion of objects with angular momentum (e.g. spin angular momentum in circularly-polarized light or orbital angular momentum in vortex beams of electrons or light). (a) Nonreciprocal effects in the motion of objects with angular momentum. The left and right situations can be linked through {**R**,**I**⊗**M**,**T**}, and the specimen constituents shown with blue (spins, magnetization) and red arrows (polarization) do have broken {**R**,**I**⊗**M**,**T**}, so can exhibit nonreciprocal directional dichroism for the motion of objects with angular momentum. (b) The motion of objects with opposite angular momenta moving in the same direction can be linked through {**M**(=**I**⊗**R**),**M**⊗**R**,**I**⊗**T**}, and all right-hand-side specimen constituents do have broken {**M**(=**I**⊗**R**),**M**⊗**R**,**I**⊗**T**}, so can exhibit the dependence on the sign of angular momentum. Not that circular polarization of light is due to its "spin" angular momentum, and vortex light beams do have "orbital" angular momentum. (c) Quasi-equilibrium process of an induced current in a *p-n* junction or a polar material under light illumination. This process varies in a systematic way under all symmetry operations, except **T**, which is not relevant for quasi-equilibrium processes. Photovoltaic effects in *p-n* junctions or bulk ferroelectrics and linear photo-galvanic effects (LPGE) correspond to this situation. (d) & (e) Quasi-equilibrium process of an induced current in a chiral or ferromagnetic material under illumination of circularly-polarized light or vortex light beam. This circular photogalvanic effect (CPGE) varies in a systematic way

under all symmetry operations, i.e. exhibits symmetry operational systematics. (f) – (h) Various Hall-effect-type transport properties on plate-shape specimens (specimen plates are in the page plane), where quasi-equilibrium processes are also involved. ***J***'s are induced charge current and thermal current under applied ***E**$_{ext}$* and *ΔT*, respectively, and (+, -) and (h: hot, c: cold) denote the induced Hall voltage and thermal gradient, respectively. The applied ***H*** in (h) is 45º away from the ***J*** direction, but is in the specimen plane.

**Nonreciprocity with Angular Momentum**

In the previous discussion on nonreciprocity, we have discoursed mostly the motion of objects without angular momentum, such as un-polarized light. The symmetry approach still works, of course, when angular momentum is added. This addition of angular momentum changes the range of the possible nonreciprocal effects. Circularly-polarized light possesses spin angular momentum. In addition, a vortex beam of light or electrons is twisted like a screw around its axis of travel, and has orbital angular momentum [39-42]. An object with angular momentum propagating, for example, along the chiral axis of a chiral material is reciprocal, since two reciprocal situations can be related through time reversal. For this reason, the optical activity of chiral materials is reciprocal, and the polarization rotation of linearly polarized light is the same for two opposite propagation directions along the chiral axis. However, when time reversal symmetry is broken in a material, the situation changes; for example, an object with angular momentum propagating along the magnetization direction is nonreciprocal, since two reciprocal situations cannot be related through any symmetry operation(s). In fact, this is the origin of the Faraday rotation in ferro-(ferri)magnetic materials, which is nonreciprocal, and these nonreciprocal Faraday effects are directly relevant to magneto-optic Kerr effects (MOKEs), i.e., the light-polarization rotation effects of linearly-polarized light when it is reflected on ferro-(ferri)magnetic surfaces (see below). In terms of symmetry, there is no difference between spin angular momentum and orbital angular momentum. The nonreciprocity of a vortex light beam with orbital angular momentum has been, for the first time, observed for THz vortex beam propagating in ferrimagnetic Tb$_3$Fe$_5$O$_{12}$ garnet [43]. The motion of an object with angular momentum along one direction can be linked with that in the opposite direction through {**R**,**I**⊗**M̲**,**T**}, and ***M*** has broken {**R**,**I**⊗**M̲**,**T**}, so ***M*** can induce nonreciprocal effect with angular momentum. Similar with ***M***, toroidal moments have also broken {**R**,**I**⊗**M̲**,**T**}, so vortex light beam with orbital angular momentum as well as circularly polarized light should exhibit nonreciprocal effects in materials with toroidal moments when propagation direction is along the toroidal moment direction. Furthermore, magnetic monopoles combined with electric field or polarization should also exhibit nonreciprocal effects with angular momentum. All of these cases are summarized in Fig. 5(a), and many of these cases have not been experimentally realized yet.

Vortex light beams with opposite orbital angular momenta or lights with opposite circular polarizations can be linked through {**M̲**(=**I**⊗**R**),**M**⊗**R**,**I**⊗**T**}, and all of the specimen constituents in Fig. 5(b) have broken {**M̲**,**M**⊗**R**,**I**⊗**T**}. Thus, these specimen constituents can exhibit the angular-momentum-sign dependence of the propagation of a vortex beam with orbital angular momentum or circularly-polarized light. The standard magnetic circular dichroism with circularly-polarized light in a transmission mode in ferro-(ferri)magnets corresponds to the angular-momentum-sign dependence in the specimen constituent with ***M*** [44]. This angular-momentum-sign dependence has been also observed for THz vortex beam with orbital angular momentum propagating in the film of ferrimagnetic Tb$_3$Fe$_5$O$_{12}$ garnet [43]. Note that ferro-rotation, which does not break space inversion, does not induce optical activity, but the ferro-rotation with external electric fields has SOS with a structural chirality, and thus induces reciprocal optical activity that is linearly proportional to external electric fields – the so-called linear electro-gyration [45].

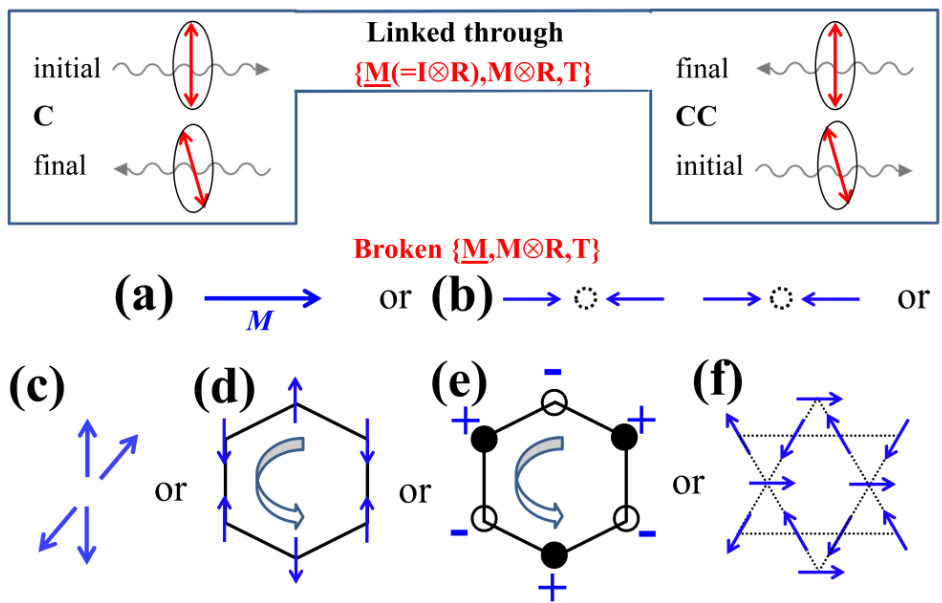

**Figure 6**, **Symmetry consideration for the polarization rotation of reflected linearly-polarized light.** Red double arrows denote the light polarization direction, and blue arrows are for spins, magnetization or magnetic field. The top-left and top-right situations can be linked through {$\underline{\mathbf{M}}$(=$\mathbf{I}\otimes\mathbf{R}$),$\mathbf{M}\otimes\mathbf{R}$,$\mathbf{T}$}, and the specimen constituents in the bottom do have broken {$\underline{\mathbf{M}}$, $\mathbf{M}\otimes\mathbf{R}$,$\mathbf{T}$}, so can exhibit the light polarization rotation (MOKE-type optical rotation). Note that light propagation for all cases is along the left-right direction, and two (buckled) honeycomb cases in (d) and (e) are rotated by 90º along the vertical axis on the page plane.

**Electronic Transport Properties with Quasi-Equilibrium Processes**

As we have discussed earlier, ***P*** in ferroelectrics, similar with the built-in electric field in *p-n* junctions, can lead to nonreciprocal electronic transport properties. In the presence of light illumination, *p-n* junctions as well as ferroelectrics can show photovoltaic effects [10,11]. This effect in a ferroelectric is called a bulk ferroelectric photovoltaic effect. Quasi-equilibrium process is involved in this photovoltaic effect, i.e. it reflects what happens when a system is quasi-equilibrated after continuous light illumination. Therefore, time reversal (**T**) operation has a limited meaning, i.e. the process of illuminating light and reaching quasi-equilibrium itself breaks time reversal symmetry. This situation is similar with the quasi-equilibrium process of electronic transport in the presence of external electric fields. In fact, the photovoltaic effect shown in Fig. 5(c) varies in a systematic manner under all symmetry operations except **T**. This symmetry operational systematics, which may be also called SOS, is powerful to understand the light illumination with angular momentum on chiral or magnetic materials. It has been reported that illumination of circularly-polarized light on chiral materials can induce photocurrent, and these effects are called circular photogalvanic effects (CPGEs) [46-48]. We can show how it works in terms of our symmetry operational systematics: a chiral material under circularly-polarized photon (or vortex light beam) fields varies in a systematic manner under all symmetry operations. For example, **M**, **I**⊗**R**, or **M**⊗**R** operation results in the situation that the simultaneous change of angular momentum sign and chirality leads to the same induced current. Interestingly,

ferro(ferri)magnetic materials or any materials in magnetic field (in the middle of Fig. 5(d)) can also exhibit CPGE since they under circularly-polarized photon (or vortex light beam) fields do exhibit symmetry operational systematics. This ferro-(ferri)magnetic CPGE seems never discussed or observed.

The concept of symmetry operational systematics of the entire experimental set-up, which combines a specimen constitution and a measuring probe/quantity, can be also applied to all Hall-effect-type transport properties where quasi-equilibrium processes are also involved [49-55]. The sets of ($E_{ext}$, +, -), ($E_{ext}$, h, c), ($\Delta T$, +, -), and ($\Delta T$, h, c) of Fig. 5(f) correspond to the Hall, Ettingshausen, Nernst, and thermal Hall effects, respectively. $E_{ext}$ in Fig. 5(g) is for the spin Hall effect and $\Delta T$ in Fig. 5(g) is for the spin Nernst (or thermal spin Hall) effect. The sets of ($E_{ext}$, +, -), ($E_{ext}$, h, c), ($\Delta T$, +, -), and ($\Delta T$, h, c) of Fig. 5(h) correspond to the planar Hall, planar Ettingshausen, planar Nernst, and planar thermal Hall effects, respectively. When the specimens have no broken symmetries, all cases with linear effects in Fig. 5(f) – (h) exhibit symmetry operational systematics, i.e. vary in systematic manners under all symmetry operations, except **T** operation. For example, the **M_**(=**I**⊗**R;** rotation axis is the vertical axis on the page plane) operation on the situation in Fig. 5(h) leads to the 180º switching of induced planar Hall voltage (+, -) or thermal gradient (h, c) by rotating **H** by 90º in the specimen plane. The linear planar Hall voltage is supposed to vary as $\sin 2\phi$, where $\phi$ is the angle between **H** and **J**. Thus, the induced planar Hall voltage (+, -) or thermal gradient (h, c) in Fig. 5(f) – (h) can be non-zero. When **H**'s in Fig. 5(f) and (h) are replaced by **M**'s, the relevant effects become "anomalous". For example, ($E_{ext}$, +, -) in Fig. 5(f) with **M**, rather than **H**, corresponds to the anomalous Hall effect. Now, when the specimens in Fig. 5(f) - (h) have certain broken symmetries, there can be non-linear effects. For example, in the case of the Hall effect, switching **H** and (+,-) simultaneously can be achieved by any of {**R**,**M_**,**M|**⊗**R•**} (the mirror of **M–** is like — (i.e. along the **J** direction and perpendicular to the page plane), the mirror of **M|** is like | (i.e. perpendicular to the **J** direction and the page plane), and the 2-fold rotation axis of **R•** (R|) is perpendicular to (vertical on) the page plane), so when a specimen with broken{**R**,**M_**,**M|**⊗**R•**}, the induced (+,-) voltage can be non-linear with H. On the other hand, switching $E_{ext}$ (**J**) and (+,-) simultaneously without changing the **H** direction in the Hall effect can be achieved by any of {**R•**,**I**}, so when a specimen with broken {**R•**,**I**} can exhibit a non-linear Hall effect in $E_{ext}$ (J). One example of a specimen with broken {**R**,**M_**,**M|**⊗**R•**} but not-broken {**R•**,**I**} is a chiral and polar sample with the chiral axis and polarization both along the direction perpendicular to the page plane, so it can show a Hall effect that is linear with $E_{ext}$ (J), but non-linear with H. Now, consider Fig. 5(f) without **H**, but with **P** along the vertical direction on the page plane. **M|** or R| operation on the experimental set-up leads to flipping $E_{ext}$/$\Delta T$ and **J** without changing **P** or induced (+,-)/(h,c), so the induced (+,-)/(h,c) should vary, for example, like $(E_{ext})^2$ or $(/\Delta T)^2$ – in other words, non-linear Hall (Ettingshausen, Nernst, thermal Hall) effects without linear effects can exist in this experimental set-up "without broken time reversal" symmetry. Emphasize that the non-reciprocal (i.e. diode-like) transport properties that we discussed in conjunction with Fig. 2 can be also understood in terms of symmetry operational systematics: all experimental set-ups, combining the specimen constituents and measuring probes/quantities, in Fig. 2 vary in systematic manners under all symmetry operations, except **T** operation.

**MOKE-type Optical Rotation**

Symmetry consideration can be also used to find the requirements for the polarization rotation of reflected linearly-polarized light with normal incidence. The top-left and top-right situations in Fig. 6 can be linked through {$\underline{\mathbf{M}}$(=$\mathbf{I}\otimes\mathbf{R}$),$\mathbf{M}\otimes\mathbf{R},\mathbf{T}$}, and all of the specimen constituents in the bottom of Fig. 6 do have broken {$\underline{\mathbf{M}}$,$\mathbf{M}\otimes\mathbf{R},\mathbf{T}$}, so can exhibit the polarization rotation. The polarization rotation of reflected light for ferromagnetic magnetization in Fig. 6(a) is the standard MOKE. Fig. 6(b) corresponds to the polarization rotation of reflected light observed in antiferromagnetic $Cr_2O_3$ without any net magnetic moment [57]. Fig 6(f) with kagome lattice is relevant to the MOKE-type optical rotation observed in antiferromagnetic $Mn_3Sn$ with a tiny net magnetic moment [58]. Note that for linearly-polarized light incident perpendicular to the kagome plane in Fig. 6(f), $\underline{\mathbf{M}}$ symmetry for the mirror along the vertical direction on the page plane and perpendicular to the kagome plane is not broken, so there exists no MOKE-type optical rotation. Fig. 6(c) corresponds to the polarization rotation of reflected light on a magnetic monopole. The specimen constituent in Fig. 6(d) is same with the honeycomb lattice with in-plane Ising antiferromagnetic order in Fig. 4(c), and the specimen constituent in Fig. 6(e) is the buckled honeycomb lattice with out-of-plane Ising antiferromagnetic order in Fig. 4(f) - emphasize that light propagation is perpendicular to the (buckled) honeycomb lattice planes. Note that the antiferromagnetic order in Fig. 6(d) on a buckled honeycomb lattice has also broken {$\underline{\mathbf{M}}$, $\mathbf{M}\otimes\mathbf{R},\mathbf{T}$}, so should exhibit the MOKE-type optical rotation. The MOKE-type optical rotations in magnetic monopoles as well as (buckled) honeycomb lattices with Ising antiferromagnetic order have not been observed yet. The magnetic monopole effect can be, in principle, observed in, for example, the A2 phase of $h$-R(Fe,Mn)$O_3$, and observing the MOKE-type optical rotation in "bulk" (buckled) honeycomb systems with Ising antiferromagnetic order requires non-cancelling contributions from different layers.

**SHG: Second Harmonic Generation**

Nonlinear optics refers to coherent optical processes where the interaction between light and matter changes the light frequency, and the simplest process of this kind is second harmonic generation (SHG) with frequency doubling. Typically, this effect is rather weak, so its observation requires high electromagnetic field strengths, as provided by a high-power pulsed laser. The point symmetry group of a compound is directly relevant to the allowed SHG, and the magnetic symmetry also matters, so SHG can be used to observe, for example, antiferromagnetic domains. For the normal-incident coherent light with the parallel arrangement of incoming and outgoing light polarizations, one can use a simple symmetry consideration for SHG – the so-called SHG(XX) with X along the light polarization direction. The left and right situations in Fig. 7(a) can be linked through {$\underline{\mathbf{R}}$,$\mathbf{M}_–$(=$\mathbf{I}\otimes\mathbf{R}|$)} (The 2-fold rotation axis of $\underline{\mathbf{R}}$ is — (i.e. along the $\mathbf{k}$ direction) and the 2-fold rotation axis of $\mathbf{R}|$ is | (i.e. vertical) on the page plane). Thus, if a material has any of {$\underline{\mathbf{R}}$,$\mathbf{M}_–$)} symmetry, then the optical response along the incoming light polarization direction (i.e., X direction) cannot have any asymmetry along the positive and negative light electric field direction, so SHG(XX) should vanish. However, any materials have broken {$\underline{\mathbf{R}}$,$\mathbf{M}_–$}, then they can exhibit SHG(XX). SHG can be utilized, for example, to induce frequency doubling or to image ferroelectric domains or antiferromagnetic domains. Herein, we briefly discuss how SHG is used to image magnetic domains. Various antiferromagnetic order configurations on (buckled) honeycomb lattice are shown in Fig. 7(c) – (f). For the 2-fold rotation ($\underline{\mathbf{R}}$) around the axis perpendicular to the page plane, honeycomb lattice has $\underline{\mathbf{R}}$ symmetry, but buckled honeycomb lattice has broken $\underline{\mathbf{R}}$ symmetry. In addition, mirror symmetry for the mirror along the horizontal direction of the buckled honeycomb lattice in Fig. 7(d) and (f) and perpendicular to the buckled honeycomb plane is broken, so SHG(XX) can exist for X perpendicular to the horizontal direction of Fig. 7(d) and (f), which can happen, for example, in the (111) plane of non-centrosymmetric (but non-polar and non-chiral) GaAs. With displayed antiferromagnetic order, R symmetry is now broken in both honeycomb lattice and buckled honeycomb lattice. Furthermore, mirror symmetry for the mirror shown with green dashed line is broken in all of the antiferromagnetic states in Fig. 7(c) – (f). Thus, SHG(XX) should exist for X perpendicular to the green dashed line. This magnetic SHG(XX) can be used to image antiferromagnetic domains, for example, in $Cr_2O_3$ [59]. Note that time reversal antiferromagnetic domains exhibit different phases of SHG response, but typically the intensity of SHG is measured experimentally, so time reversal domains do not show any contrast even though time reversal domain "walls" can show contrast due to destructive interference at domain walls. However, when magnetic SHG is combined with crystallographic SHG, the time reversal domains can show contrast since magnetic SHG's of time reversal domains are added to crystallographic SHG with opposite signs [59].

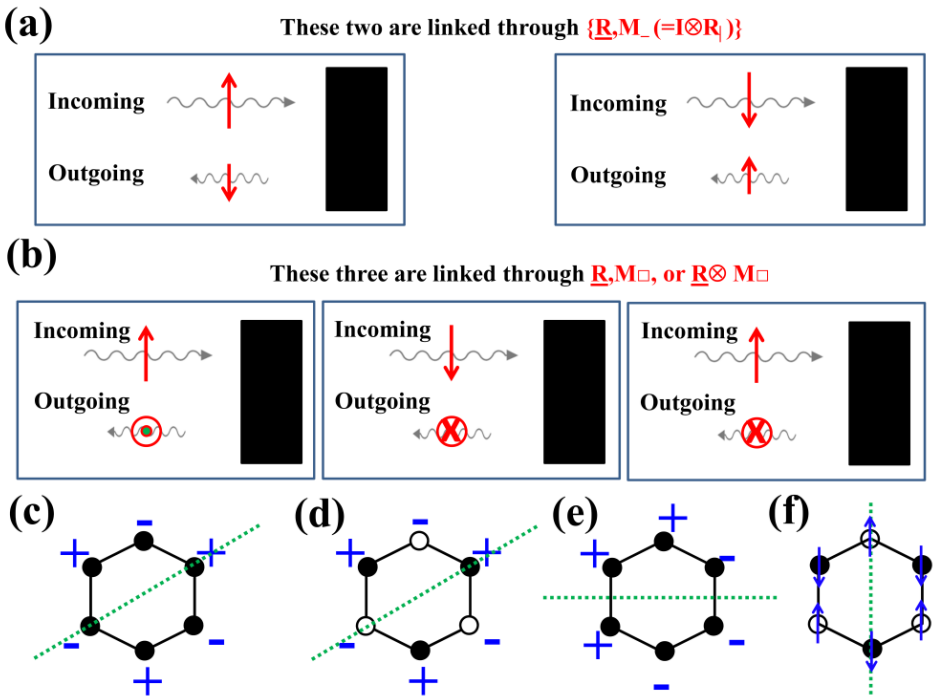

**Figure 7**, **Symmetry consideration for SHG with normal-incident coherent light.** (a) The parallel arrangement of incoming and outgoing light polarizations, relevant to SHG(XX). The left and right situations in Fig. 6(a) can be linked through {R,M− (=I⊗R|)}, and SHG can occur when {R,M−} are broken. Red arrows denote coherent light electric field (polarization). (b) The perpendicular arrangement of incoming and outgoing light polarizations, relevant to SHG(XY). The 1st and 2nd situations can be linked through R, and the 1st and 3rd situation can be linked through M□, and SHG(XY) can occur when {R,M□} are broken. Red arrows denote coherent light electric field (polarization). (c) and (e) Various out-of-plane antiferromagnetic order configurations on honeycomb lattice. (d) and (f) Various antiferromagnetic order configurations on buckled honeycomb lattice with spins below (solid circles) and above (open circles) the page plane. The antiferromagnetic lattices in (c) – (f) for coherent light incident perpendicular to the 2D lattice have broken R, and have also broken M− or M□ (green dashed line, perpendicular to the page

plane). Thus, SHG(XX) can occur in these situations for X along the direction perpendicular to the green mirror and SHG(XY) with Y along the direction perpendicular to the green mirror, which can be utilized to visualize antiferromagnetic domains.

The 1$^{st}$ and 2$^{nd}$ situations in Fig. 7(b) can be linked through **R**, and the 1$^{st}$ and 3$^{rd}$ situations in Fig. 7(b) can be linked through **M**$_\square$. (the mirror of **M**$_\square$ is like □ on the page plane). Thus, if a material has any of {**R**,**M**$_\square$} symmetry, then the optical response along the direction (i.e., Y direction) perpendicular to both the incoming light polarization X direction and light propagation direction cannot have any asymmetry along the positive and negative Y directions, so SHG(XY) should vanish. However, SHG(XY) can occur when {**R**,**M**$_\square$} are broken. Now, we make the following powerful statement for SHG with normal-incident coherent light: [1] When **R** symmetry is broken, then there exists, at least, one broken **M**$_n$ (mirror normal to the probing surface). [2] SHG(XX) can be non-zero when X is perpendicular to **M**$_n$ and SHG(XY) can be non-zero when X is parallel to **M**$_n$ (i.e. Y is perpendicular to **M**$_n$). [2] has been already discussed, and [1] can be proven like this: If there is no broken **M**$_n$, then we can choose two perpendicular **M**$_{n1}$ and **M**$_{n2}$ with no-broken mirror symmetry. Then, **M**$_{n1}$⊗**M**$_{n2}$=(I⊗**R**$_1$)⊗(I⊗**R**$_2$)=**R**$_1$⊗**R**$_2$ (the rotation axes of **R**$_1$ and **R**$_2$ are perpendicular to **M**$_{n1}$ and **M**$_{n2}$, respectively). Now, we have **R**$_1$⊗**R**$_2$⊗**R**=**1, so R**$_1$⊗**R**$_2$=**R**$^{-1}$, which is same with **R**. Thus, **M**$_{n1}$⊗**M**$_{n2}$=**R**$_1$⊗**R**$_2$=**R**$^{-1}$=**R**, so **R** symmetry is not broken. The proof is complete.

In summary, we have discussed that the occurrence of certain physical phenomena can be understood in terms of SOS between specimen constituents and measuring probes/quantities. The SOS approach can be applied to numerous physical phenomena such as nonreciprocity, magnetism-induced ferroelectricity, linear magnetoelectricity, optical activities (including the Faraday and magneto-optic Kerr rotation), photogalvanic effects, and second harmonic generation, and can be leveraged to identify new materials that potentially exhibit the desired physical phenomena. Note that the approach with (magnetic) point symmetry groups to understand numerous physical phenomena is standard and can often produce the complete picture. However, we frequently we do not know the exact point symmetry groups of new materials, especially magnetic point symmetry groups, and understanding high-rank tensorial characteristics of certain observables in complex (magnetic) point symmetry groups can be tedious and complicated. This standard approach can readily miss specimen constituents that should be able to exhibit certain physical phenomena, many of which we have deliberated. We have discussed various specimen constituents, many of which are quasi-1D, and the 2D cases of (buckled) honeycomb or kagome layers. Evidently, in order to result in bulk effects, the contributions from different chains or layers have to be constructively added in real 3D materials. Our SOS approach can readily lead to numerous new predictions that invite experimental confirmation. The intriguing examples include [1] magnetic helicity has SOS with structural chirality, so may exhibit optical activity, [2] transverse magneto-chiral effects in optical transmission or electronic transport in Fig. 1(c) can occur in mono-axial chiral materials, [3] toroidal moment with rotating spins in Fig. 1(g) should show nonreciprocity in the propagation of unpolarized light or spin wave, [4] Neel-type ferroelectric wall motion in the direction perpendicular to the wall can produce magnetization along the direction perpendicular to the polarization rotation plane, [5] many cases in Fig. 4 and Fig. 5(a) and (b) have never been observed/considered, but can be realized in real materials, [6] ferro-(ferri)magnetic CPGE in Fig. 5(c) has never been observed/considered, but can be observed in real materials, [7] a number of non-linear Hall-effect-type transport properties discussed in conjunction with Fig. 5(f) have not been observed/considered, but can be readily experimentally tested, and [8] a MOKE-type optical rotation can be observed in magnetic monopole systems and (buckled) honeycomb systems with Ising antiferromagnetism in the bottom of Fig. 6, and [9] SHG due to antiferromagnetic order on (buckled) honeycomb lattice in Fig. 7(c), (e), (f) and (g), has never observed, and needs to be experimentally verified.

We have unveiled the following important rule for SHG for normal-incident coherent light: *When **R** symmetry is broken, then there is, at least, one broken $\mathbf{M_n}$ (mirror normal to the probing surface). SHG(XX) can be non-zero when X is perpendicular to $\mathbf{M_n}$ and SHG(XY) can be non-zero when X is parallel to $\mathbf{M_n}$ (i.e. Y is perpendicular to $\mathbf{M_n}$).* This statement works for crystallographic structures as well as magnetic structures. One further challenge in terms of measurements is spatially(<1 µm)-resolved imaging of antiferromagnetic (including magnetic monopolar or toroidal) or ferro-rotational domains/domain walls using, for example, nonreciprocity or optical polarization rotation. Vortex light or electron beams with orbital angular momentum appear to be powerful new tools to explore new physical phenomena associated with broken symmetries. We emphasize that the SOS arguments do not tell neither the microscopic mechanism for physical phenomena nor their magnitudes. However, it turns out that, for example, all magnets with cycloidal spin order, corresponding to Fig. 3(a), always show experimentally-measurable *P*. Thus, these situations appear to resonate with the famous statement of Murray Gell-Mann: "everything not forbidden is compulsory" [60]. In other words, when certain physical phenomenon is allowed by the consideration of broken symmetries, the effect is often large enough, so experimentally observable.

**Methods:** The relationships of specimen constituents (i.e., lattice distortions or spin arrangements, in external fields or other environments, etc.) and one-dimensional measuring probes/quantities (i.e., propagating light, electrons or other particles in various polarization states, including vortex beams of light and electrons, bulk polarization or magnetization, etc.) are analyzed in term of the characteristics under various symmetry operations of rotation, space inversion, mirror reflection, and time reversal. When specimen constituents and measuring probes/quantities share the same broken symmetries, except translation symmetry along the measuring direction, they are said to exhibit symmetry operation similarities (SOS), and the corresponding phenomena can occur. We have also considered all symmetry operations linking two conjugate measuring probes/quantities such as quasi-particles with angular momentum moving two opposite directions, propagating quasiparticles with two opposite signs of angular momentum, the reflection of linearly-polarized light with two opposite MOKE-type optical rotations, and non-linear response (second harmonic generation-type) of coherent normal-incident light with two opposite light polarization directions. If all symmetries of those linking symmetry operations are broken in a specimen constitution, then it can exhibit the corresponding phenomenon. For electronic transport properties, including Hall-effect-type transport effects, where quasi-equilibrium processes are involved, we ignore time reversal symmetry, and it is more convenient to consider the entire experimental set-ups, which combine specimen constituents and measuring probes/quantities. When an experimental set-up varies systematically under all symmetry operations, except time reversal operation, then it is said to exhibit symmetry operational systematics, and the corresponding phenomenon can arise.

**Acknowledgement**: The work was supported by the DOE under Grant No. DOE: DE-FG02-07ER46382 and the Gordon and Betty Moore Foundation's EPiQS Initiative through Grant GBMF4413 to the Rutgers Center for Emergent Materials. The author has greatly benefited by discussion with Manfred Fiebig, FeiTing Huang, Valery Kiryukhin, Roman Pisarev, Diyar Talbayev, David Vanderbilt, and Liang Wu. A significant part of SHG section results from insightful discussions with Liuyan Zhao.

**Data availability**: The datasets generated during the current study are available from the corresponding author on reasonable request.

**Competing Interests**: The authors declare no conflict of interest.